\documentclass[10pt]{article}
\title{{\bf Construction of gauge invariant effective\\
nucleonic theories: functional approach }}
\author{ \\
\\
\\
\\ H. W. L. Naus 
\\
\\
 {\em Institute for Theoretical Physics, University of Hannover}\\
{\em Appelstr. 2, 30167 Hannover, Germany}\\
 \\
\\
\\
\\}
\date{\today \
\\
 \begin{abstract}
Starting from relativistic quantum field theories, describing
interacting nucleons and pions coupled to the dynamical 
electromagnetic field, the pion degrees of freedom are eliminated
by means of functional integration. Apart from taking into account
some operators perturbatively in $e$, e.g. the vacuum polarization,
this procedure is exact, giving effective theories for nucleons and photons.
The subsequent nonrelativistic reduction yields the corresponding
nonrelativistic quantum field theory. The latter is unique, irrespective
of the precize form of the original nucleon-pion interaction.
Nucleonic potentials and electromagnetic interactions are mutually
consistent. Local gauge invariance is satisfied at any stage of the
formal developments. 
\end{abstract}}
\begin{document}
\baselineskip=12.0pt
\maketitle
\setcounter{equation}{0}
\pagebreak
\section{Introduction}
Low energy strong interaction physics can be rather well described in terms
of effective theories with phenomenological interactions. Coupling the
electromagnetic field, however, is ambiguous since it is essentially
impossible to uniquely construct the relevant currents without a 
deeper understanding of the origin of the interactions. Local gauge invariance
clearly shows the necessity of additional, {\it i.e.}, beyond the
standard one-body terms, electromagnetic interaction terms by means of
constraints. These gauge conditions for `exchange currents'
cannot fix them completely: transverse part of currents are  
not constrained. It may appear frustrating that these eventually
yield the physical amplitudes.
Nevertheless, in particular in combination with Lorentz- or rotational
invariance, local gauge invariance is a powerful concept in effective
models also.

Electromagnetic current conservation
is a consequence of global $U(1)$ gauge symmetry. Therefore, the
question arises whether the stronger condition of
local gauge invariance is necessary in nonrelativistic theories.
This difference, global versus local, may have practical consequences,
for example with respect to Siegert's hypothesis \cite{Sie}.
The latter can be shown  to be a consequence of local
$U(1)$ gauge symmetry in nonrelativistic quantum mechanics \cite{Naus}.
Thus effective theories where the charge density is modified, e.g.
\cite{Hy}, are not invariant under the usual local phase transformations
although the current is conserved.
The breaking  seems to be caused by the choosen
procedure of eliminating degrees of freedom by means 
of unitary transformations followed by projection on a subspace
of the Hilbert space \cite{FSK}. 
A more recent example of this method
is the derivation of an effective meson-exchange model
for pion-nucleon scattering and pion photo- and electroproduction \cite{Sato}. 

Another method to obtain  nuclear Hamiltonians
and effective electromagnetic currents from a model with
interacting nucleons and mesons is the (extended) $S$-matrix approach
\cite{Adam}. Alternatively, Friar developed a perturbation technique
to eliminate meson degrees of freedom exploiting the equations
of motion \cite{Friar}.
Thus, effective baryonic theories are often perturbatively constructed
via the elimination of mesons from a interacting model.
In this work we address the following problem of this kind:
how to get, from a pion-nucleon Lagrangian, to an effective
nucleon theory including electromagnetic interactions.
In contrast to the examples above, the elimination procedure
is nonperturbative.
In principle our methods are general and can be applied
to other models as well; in practice the feasibility depends on the appearing
interactions. 
It is essential that we couple the electromagnetic field
from the very beginning and take this gauge invariant Lagrangian
as starting point. Then we derive 
a relativistic quantum field theory describing  only nucleons and 
photons. The corresponding Lagrangian contains nonlocal interactions.
Eventually we arrive at a theory with nonrelativistic nucleons.
At any stage local gauge invariance is satisfied and, consequently,
Ward-Takahashi identities \cite{WT} hold.
In this way, given the original action, the deduced interactions and
electromagnetic currents are mutually consistent and no ambiguities arise.

It should be emphasized that our starting point is a local
field theory, describing point-like hadrons. No form factors
are put in by hand; the structure of the nucleon is generated
by the pions. Since renormalizability is not an issue here, one may of course
put in a Pauli-term, which partly accounts for the anomalous
magnetic moment of the nucleon. As is well-known, such a term
is separately gauge invariant and obviously does not
contain the pion field. Consequently, the developments and
discussions in this work do not depend on its possible presence. 

We work in the path-integral formulation of quantum field theory.
The explicit elimination of the pions is done by means of
functional integration. We have choosen this approach for several 
reasons. First, it allows for an `exact', nonperturbative
removal of the pion fields. Secondly, the electromagnetic field
can be treated dynamically. However, the effective self-coupling
of the photons -induced by charged pions- as well as 
some appearing operators are only calculated perturbatively in $e$.
Thirdly, it is relatively easy to demonstrate that the appearing
effective actions reflect the local $U(1)$ gauge symmetry.
Finally, by means of sources one can easily identify relevant
Green functions, including those with an external pion.
In other words, even after integrating out the pions one could
still calculate, e.g., pion photoproduction.

%This paper is organized as follows.
In the next section
we present the formalism and subsequently
derive the effective actions for nucleons interacting with
an external electromagnetic field, starting 
from pseudoscalar, pseudovector and mixed pion-nucleon couplings. 
Section 3 contains the extension to dynamical
photons. Local gauge invariance is explicitly verified
in section 4. Section 5 deals with nonrelativistic
approximations. Finally, we summarize and
discuss possible applications in section 6.
Some technical aspects are relegated to Appendices.

\section{Functional integration}
\subsection{Path-integral quantization}
Let us first address the notation and some well-known aspects
of the formalism. Since we want to concentrate on gauge invariance
aspects, we do not explicitly introduce isospin. Recall that
the electromagnetic interaction breaks isospin symmetry.
Our choices of the strong interaction terms, however, respect
this symmetry. The proton and neutron fields are denoted
by $\Psi_p, \Psi_n$, respectively. The neutral pion, $\pi^0$, is
decribed with a real scalar field $\phi$. For the charged pions,
$\pi^+$ and $\pi^-$, we introduce one complex scalar field $\Phi$.
The relation to the real isovector components is given by
$\Phi = \frac{1}{2} \sqrt{2} (\phi_1 + i \phi_2) $. We will use
pseudoscalar as well pseudovector
pion-nucleon interactions; the corresponding coupling constants
$g$ and $f$ are related: $\frac{f}{m} = \frac{g}{2M}$, with $m$ the (common)
pion mass and $M$ the (common) nucleon mass. The electromagnetic
field strength tensor $F_{\mu \nu}$ is given in terms of the gauge
fields $A_\mu$ by 
$F_{\mu \nu} = \partial_\mu  A_\nu - \partial_\nu  A_\mu$. 

In the path-integral formulation of quantum field theory the
generating functional $Z$ is the central object from which
Green's functions and $S$-matrix elements (in principle) can
be obtained. Of course, the generating functional is 
determined by the Lagrangian of the theory. The general formalism
is presented in modern books on field theory, e.g. \cite{Ry}.
Here we briefly
sketch the connection for the models under consideration.

The Lagrangian density ${\cal L}(x)$ -to be specified later- can be 
extended with sources and gauge-fixing term for the Lorentz-gauge
\begin{equation}
{\cal L}_s(x) = {\cal L} (x) + \bar{\eta}_p \Psi_p +\bar{\Psi}_p \eta_p
 + \bar{\eta}_n \Psi_n +\bar{\Psi}_n \eta_n 
+\Phi^* J + J^* \Phi + \phi J^0 
+{\cal J}^{\mu} A_{\mu} - \frac{1}{2\alpha} (\partial_{\mu} A^{\mu})^2.
\label{eq:sou}
\end{equation}
The (source dependent) action $S_s$
%\begin{equation}
%S_s = \int d^4 x \, {\cal L}_s (x) ,
%\label{eq:actso}
%\end{equation}
appears in the generating functional $Z$,
\begin{equation}
Z = {\cal N} \int 
{\cal D} A_{\mu} {\cal D}\bar{\Psi}_{p} {\cal D}\Psi_{p} 
{\cal D}\bar{\Psi}_{n} {\cal D}\Psi_{n} {\cal D}\Phi^* 
{\cal D}\Phi {\cal D}\phi \exp(iS_s) = 
{\cal N} \int {\cal D} A_\mu \cdots {\cal D} \phi
\exp\left(i \int d^4 x \, {\cal L}_s (x) \right) \, .
\label{eq:Zsou}
\end{equation}
The Faddeev-Popov ghost term has been absorbed in the
the normalization ${\cal N}$; this is possible because in abelian theories
the ghosts do not couple
to physical fields \cite{Ry}.  It should be remarked that
for the formal developments
in this section it is irrelevant whether or not one includes the
gauge-fixing term, fermion sources and the photon source.  In other words,
these terms can also be added in a later stadium, in particular after
integrating out the pion fields. 

For the theories we will consider below, the action without the
sources and the gauge-fixing term is locally gauge invariant.
The local gauge transformations are explicitly given by
\begin{eqnarray}
A_\mu &\rightarrow& A_\mu + \partial_\mu \chi \, , \nonumber \\
\Psi_p &\rightarrow& \exp(-ie\chi) \Psi_p \, , \hskip 2cm
%\bar{\Psi}_p \rightarrow \exp(ie\chi) \bar{\Psi}_p \, , \nonumber \\
\Phi \rightarrow \exp(-ie\chi) \Phi \, ,
%\hskip 2.2cm
%\Phi^* \rightarrow \exp(ie\chi) \Phi^* \,  ,
\end{eqnarray}
where $\chi (x)$ is an arbitrary function. The fields $\Psi_n$ and $\phi$
are invariant since they describe neutral particles.
This gauge symmetry leads to relations between vertex functions and
propagators, {\it i.e.}, Ward-Takahashi identities. In the functional
formalism, they are readily derived starting from  the
generating functional $Z$ \cite{Ry}. Note that the integration measure in
$Z$ is also invariant. Therefore one may {\it a priori} expect
that exactly integrating  out the pions yields an effective action
which, again without sources and gauge fixing term, is invariant
under gauge transformations of the fields left. In turn,
this implies Ward-Takahashi identities for the relevant Green's
functions. Nevertheless, it seems to be appropriate to
explicitly check the invariance of the appearing effective actions,
especially if further approximations are involved.

\subsection{Pseudoscalar coupling}
The Lagrangian density for nucleons and pions with pseudoscalar
coupling interacting with the dynamical electromagnetic field reads
\begin{eqnarray}
{\cal L}(x) &=&  i \bar{\Psi}_p \gamma^\mu (\partial_\mu + ie A_\mu) \Psi_p
+  i \bar{\Psi}_n \gamma^\mu \partial_\mu  \Psi_n -M \bar{\Psi}_p \Psi_p
-M \bar{\Psi}_n \Psi_n -\frac{1}{4} F_{\mu \nu} F^{\mu \nu} \nonumber \\
&+& (\partial_\mu - ie A_\mu) \Phi^* (\partial^\mu + ie A^\mu) \Phi
+\frac{1}{2} \partial_\mu  \phi \partial^\mu \phi 
-m^2\Phi^* \Phi -\frac{1}{2}m^2 \phi^2 \nonumber \\
&-& i g \sqrt{2}  \bar{\Psi}_p \gamma_5 \Psi_n \Phi
- i g \sqrt{2}  \bar{\Psi}_n \gamma_5 \Psi_p \Phi^*
- i g (  \bar{\Psi}_p \gamma_5 \Psi_p - \bar{\Psi}_n \gamma_5 \Psi_n ) \phi.
\end{eqnarray}
As mentioned in the introduction, one can add Pauli terms for the nucleons;
they are proportional to $ \bar{\Psi} \sigma^{\mu \nu} \Psi F_{\mu \nu}$.
The action $S= \int d^4 x \, {\cal L} (x) $
is locally gauge invariant in either case.
In order to prepare the functional integration we rewrite
the terms where the covariant derivative acts on the complex field,
\begin{equation}
\int d^4 x \,  [ (\partial_\mu - ie A_\mu) \Phi^* (\partial^\mu + ie A^\mu) \Phi
-m^2 \Phi^* \Phi ]
= - \int d^4 x \, \Phi^* {\cal O}_A \Phi  \; ,
\end{equation}
with the differential operator
\begin{equation}
{\cal O}_A  = \partial_\mu \partial^\mu + m^2 
+ 2ie A_{\mu} \partial^\mu 
+ie (\partial^\mu A_\mu) -e^2 A^2 \;.
\end{equation}
In the following we will also use the operator ${\cal O}$
which can be obtained from ${\cal O}_A$ by putting
$e = 0$.
Now we extend this action by the inclusion of sources and
gauge fixing term (cf. eqs. (\ref{eq:sou}, \ref{eq:Zsou})), and 
rearrange the Lagrangian as follows
\begin{equation}
{\cal L}_s = {\cal L}_1 - \Phi^* {\cal O}_A \Phi + \Phi^* F
+ \tilde{F} \Phi -\frac{1}{2} \phi {\cal O} \phi +\phi F_0 \, ,
\end{equation}
with the nucleon-photon Lagrangian
\begin{eqnarray}
{\cal L}_1(x) &=&  i \bar{\Psi}_p \gamma^\mu (\partial_\mu + ie A_\mu) \Psi_p
+  i \bar{\Psi}_n \gamma^\mu \partial_\mu  \Psi_n -M \bar{\Psi}_p \Psi_p
-M \bar{\Psi}_n \Psi_n -\frac{1}{4} F_{\mu \nu} F^{\mu \nu} \nonumber \\
 &+&  \bar{\eta}_p \Psi_p +\bar{\Psi}_p \eta_p
 + \bar{\eta}_n \Psi_n +\bar{\Psi}_n \eta_n 
+{\cal J}^{\mu} A_{\mu} - \frac{1}{2\alpha} (\partial_{\mu} A^{\mu})^2 \, ,
\label{eq:npho}
\end{eqnarray}
and the generalized sources
\begin{eqnarray}
F &=& - i g \sqrt{2} \bar{\Psi}_n \gamma_5 \Psi_p + J \, , \nonumber \\
\tilde{F} &=& - i g \sqrt{2} \bar{\Psi}_p \gamma_5 \Psi_n + J^*  \, ,
\nonumber \\
F_0 &=& - i g ( \bar{\Psi}_p \gamma_5 \Psi_p  -
  \bar{\Psi}_n \gamma_5 \Psi_n) + J_0 \, .
\end{eqnarray}
The latter effectively account for the pion-nucleon interaction.

We define the effective action $S_{ef}$ by 
functional integration over the pion fields,
\begin{eqnarray}
\exp \left(iS_{ef}\right) &=& {\cal N} \int {\cal D} \Phi^*
{\cal D} \Phi {\cal D} \phi \exp \left(iS_s\right)  \nonumber \\
  &=& {\cal N} \exp\left(i \int d^4 x \, {\cal L}_1(x)\right) 
 \cdot  \int {\cal D} \phi \exp\left(-i \int d^4 x \, [
\frac{1}{2} \phi {\cal O} \phi -\phi F_0 ] \right) \nonumber \\
 & \cdot & \int {\cal D} \Phi^* {\cal D} \Phi \exp\left(-i \int d^4 x \, [
\Phi^* {\cal O}_A \Phi -\Phi^* F - \tilde{F} \Phi ] \right) \, .
\end{eqnarray}
Since the generalized sources contain the fermion fields, this
integral yields effective fermion-fermion interactions.
The exchange currents are generated by the operator ${\cal O}_A$,
which explicitly depends on the gauge field. Despite these
nontrivial dependences the pion fields can now readily be
integrated out; the appearing integrals are essentially
gaussian.

Let us start with the neutral pions. The functional integral
gives
\begin{equation}
  \int {\cal D} \phi \exp\left(-i \int d^4 x \, [
\frac{1}{2} \phi {\cal O} \phi -\phi F_0 ] \right) = 
C \exp\left( \frac{-i}{2} \int d^4 x \, \int d^4 y \, 
F_0(x) G_0(x-y) F_0(y) \right) \, ,
\end{equation}
where $C \propto (\det {\cal O})^{-\frac{1}{2}}$
is an infinite constant, to be absorbed in ${\cal N}$.
Furthermore we recognize the Feynman propagator $G_0$,
which is defined as the inverse of the differential
operator ${\cal O}$,
\begin{equation}
{\cal O}(x) G_0(x-y) = -\delta^4(x-y) \, ,
\label{eq:fprod}
\end{equation}
and explicitly reads
\begin{equation}
G_0(z) = \int \frac{d^4 k}{(2 \pi)^4}  \frac{\exp(ikz)}{k^2-m^2+i\epsilon}  \, .
\label{eq:fpro}
\end{equation}
Note that we (re-)inserted the $i\epsilon$ prescription.
In this way we obtain the following nonlocal term in the 
effective Lagrangian density
\begin{equation}
\Delta{\cal L}_0(x) = - \frac{1}{2}  \int d^4 y \, 
F_0(x) G_0(x-y) F_0(y) \, .
\label{eq:L0}
\end{equation}
Apart from the four-fermion interaction, it describes the
coupling of the $\pi^0$ sources.

The integration over the complex fields is more involved.
The reason is the $A$-dependence of the differential
operator ${\cal O}_A$. It implies that, for dynamical photon
fields, its determinant cannot be absorbed in the normalization.  
Therefore, this determinant has to be explicitly evaluated.
We postpone this issue and first restrict ourselves to
external electromagnetic fields. Thus we do not consider here
the full nucleon-photon  Lagrangian ${\cal L}_1$,
but rather
\begin{equation}
{\cal L}^{ex}_1(x) =
{\cal L}_1(x) + \frac{1}{4} F_{\mu \nu} F^{\mu \nu}
-{\cal J}^{\mu} A_{\mu} + \frac{1}{2\alpha} (\partial_{\mu} A^{\mu})^2 \, .
\label{eq:nphoeff}
\end{equation}
Then the formal $\Phi, \Phi^*$ integration is again straightforward
\begin{equation}
  \int {\cal D} \Phi^* {\cal D} \Phi
\exp(-i \int d^4 x \, [
\Phi^* {\cal O}_A  \Phi -\Phi^* F - \tilde{F} \Phi] ) = 
D \exp( -i \int d^4 x \, \int d^4 y \, 
\tilde{F}(x) G_A(x, y) F(y) ) \, ,
\end{equation}
where $G_A$ satisfies
\begin{equation}
{\cal O}_A(x) G_A(x,y) = -\delta^4(x-y) \, .
\label{eq:OA}
\end{equation}
This result corresponds to the  nonlocal term
\begin{equation}
\Delta{\cal L}_A(x) =
- \int d^4 y \, \tilde{F}(x) G_A(x, y) F(y) \, ,
\label{eq:LA}
\end{equation}
in the effective Lagrangian density.  The latter is given by
\begin{equation}
{\cal L}^{ef, ex} (x) = {\cal L}_1^{ex}(x) 
+ \Delta{\cal L}_0(x) 
+ \Delta{\cal L}_A(x) .
\label{eq:Leff}
\end{equation}

The construction of the effective action for
pseudoscalar pion-nucleon coupling and external electromagnetic
fields is complete at this point. However, the quantity $G_A$,
describing the propagation of charged pions in a (given) electromagnetic
field, has only been
defined via a differential equation and -in contrast to $G_0$-
has not been explicitly given yet. For general $A$-fields we
were not able to construct an exact solution in a closed form. Nevertheless,
one can calculate $G_A$ perturbatively in the electromagnetic
coupling $e$. Note that the operator ${\cal O}_A$
actually contains first and second order terms:
\begin{equation}
{\cal O}_A = {\cal O} + e {\cal D}  -e^2 A^2 \, , 
\label{eq:Oper}
\end{equation}
with
\begin{equation}
{\cal D} = 2i A_\mu \partial^\mu + i (\partial^\mu A_\mu) \, .
\label{eq:Dper}
\end{equation}
In Appendix A we will derive a method to construct
$G_A$ to any order in the charge  $e$. 
Herewith, the electromagnetic interactions of
the nucleons can also be determined to arbitrary order in $e$.

\subsection{Pseudovector coupling}
Let us now consider pseudovector pion-nucleon coupling.
In this case the interaction contains a derivative of
the pion field.
For example, instead of the pseudoscalar term $\sqrt{2} ig
\bar{\Psi}_p \gamma_5 \Psi_n \Phi$ one has the pseudovector
term $\frac{\sqrt{2} f}{m} 
\bar{\Psi}_p \gamma_5 \gamma^\mu (\partial_\mu \Phi) \Psi_n$. 
Since the derivatives also act on the complex fields, describing
charged pions, minimal coupling of the electromagnetic
fields generates additional electromagnetic interactions. 
These are the well-known `contact terms'; indeed they are
required by local gauge invariance. The full pseudovector Lagrangian
density reads
\begin{eqnarray}
{\cal L}^{PV}(x) &=&  i \bar{\Psi}_p \gamma^\mu (\partial_\mu + ie A_\mu) \Psi_p
+  i \bar{\Psi}_n \gamma^\mu \partial_\mu  \Psi_n -M \bar{\Psi}_p \Psi_p
-M \bar{\Psi}_n \Psi_n -\frac{1}{4} F_{\mu \nu} F^{\mu \nu} \nonumber \\
&+& (\partial_\mu - ie A_\mu) \Phi^* (\partial^\mu + ie A^\mu) \Phi
+\frac{1}{2} \partial_\mu  \phi \partial^\mu \phi 
-m^2\Phi^* \Phi -\frac{1}{2}m^2 \phi^2 \nonumber \\
&-& \frac{\sqrt{2} f}{m}  \bar{\Psi}_p \gamma_5 
\gamma_\mu \left[ (\partial^\mu +ie A^\mu) \Phi \right] \Psi_n 
- \frac{\sqrt{2}f}{m}  \bar{\Psi}_n \gamma_5 
\gamma_\mu \left[ (\partial^\mu -ie A^\mu) \Phi^* \right] \Psi_p  \nonumber \\
&-& \frac{f}{m} \left[ 
\bar{\Psi}_p \gamma_5 \gamma_\mu (\partial^\mu \phi) \Psi_p -
\bar{\Psi}_n \gamma_5 \gamma_\mu (\partial^\mu \phi) \Psi_n \right] \,.
\end{eqnarray}
Analogous to the pseudoscalar case, we consider the action
including sources and gauge fixing term and rewrite $S^{PV}$ in terms
of the operators ${\cal O}_A$ and ${\cal O}$. Furthermore, after
more integrations by parts, one can define the pseudovector
generalized sources as
\begin{eqnarray}
F_A^{PV} &=& \frac{\sqrt{2}f}{m} \left[ \bar{\Psi}_n \gamma_5 \gamma_\mu
(\partial^\mu  \Psi_p) + (\partial^\mu \bar{\Psi}_n) \gamma_5 \gamma_\mu
\Psi_p +ie \bar{\Psi}_n \gamma_5 \gamma_\mu A^\mu \Psi_p
\right] + J \, , \nonumber \\
\tilde{F}_A^{PV} &=& \frac{\sqrt{2}f}{m} \left[ \bar{\Psi}_p \gamma_5 \gamma_\mu
(\partial^\mu  \Psi_n) + (\partial^\mu \bar{\Psi}_p) \gamma_5 \gamma_\mu
\Psi_n -ie \bar{\Psi}_p \gamma_5 \gamma_\mu A^\mu \Psi_n
\right] + J^*  \, , \nonumber \\
F_0^{PV} &=& \frac{f}{m} \left[ \bar{\Psi}_p \gamma_5 \gamma_\mu (\partial^\mu
 \Psi_p) + (\partial^\mu \bar{\Psi}_p) \gamma_5 \gamma_\mu \Psi_p 
- \bar{\Psi}_n \gamma_5 \gamma_\mu (\partial^\mu
 \Psi_n) - (\partial^\mu \bar{\Psi}_n) \gamma_5 \gamma_\mu \Psi_n 
\right] + J_0  \, . 
\end{eqnarray}
In contrast to pseudoscalar coupling, these generalized
sources explicitly depend on the gauge field.  The source
dependent Lagrangian can now be written as
\begin{equation}
{\cal L}^{PV}_s = {\cal L}_1 - \Phi^* {\cal O}_A \Phi + \Phi^* F^{PV}_A
+ \tilde{F}^{PV}_A \Phi -\frac{1}{2} \phi {\cal O} \phi +\phi F_0^{PV} \, .
\end{equation}
We see that pseudovector versus pseudoscalar
coupling amounts to redefining the generalized sources.

Thus, the simple gaussian structure allowing for an
exact integration over the pion fields is also present
in this case. The complexity of the generalized
sources is irrelevant at this point.
Again, we first treat the electromagnetic field
as external. This means that we replace ${\cal L}_1$ by
${\cal L}_1^{ex}$ and absorb the determinant of the
operator ${\cal O}_A$ into the normalization ${\cal N}$. 
The formal results of the functional integration can immediately
be read off from eqs. (\ref{eq:L0}, \ref{eq:LA}, \ref{eq:Leff}).
We obtain the effective Lagrangian
\begin{equation}
{\cal L}^{ef, ex}_{PV} (x) = {\cal L}_1^{ex}(x) 
+ \Delta{\cal L}_0^{PV}(x) 
+ \Delta{\cal L}_A^{PV}(x) \, ,
\label{eq:LeffPV}
\end{equation}
with
\begin{equation}
\Delta{\cal L}_0^{PV}(x) = - \frac{1}{2}  \int d^4 y \, 
F_0^{PV}(x) G_0(x-y) F_0^{PV}(y) \, ,
\end{equation}
and
\begin{equation}
\Delta{\cal L}_A^{PV}(x) =
- \int d^4 y \, \tilde{F}_A^{PV}(x) G_A(x, y) F_A^{PV}(y) \, .
\end{equation}
Though the elimination of the pions was completely
analogous, the resulting effective interactions are of course different for
pseudoscalar and pseudovector coupling. Concomitantly, the
additional electromagnetic interactions are consistently generated.
Exchange currents are not only contained in $G_A$, due to the coupling of
the photon to a propagating
charged pion, but also in $F_A^{PV}$ in the form of contact terms.

\subsection{Mixed coupling}
With the apparatus developed so far, the extension to
mixed, {\it i.e.}, pseudoscalar and pseudovector pion-nucleon
terms is straightforward. We will take as Lagrangian density
\begin{eqnarray}
{\cal L}(x) &=&  i \bar{\Psi}_p \gamma^\mu (\partial_\mu + ie A_\mu) \Psi_p
+  i \bar{\Psi}_n \gamma^\mu \partial_\mu  \Psi_n -M \bar{\Psi}_p \Psi_p
-M \bar{\Psi}_n \Psi_n -\frac{1}{4} F_{\mu \nu} F^{\mu \nu} \nonumber \\
&+& (\partial_\mu - ie A_\mu) \Phi^* (\partial^\mu + ie A^\mu) \Phi
+\frac{1}{2} \partial_\mu  \phi \partial^\mu \phi 
-m^2\Phi^* \Phi -\frac{1}{2}m^2 \phi^2 \nonumber \\
&-& i g \beta_1 \left[ \sqrt{2}  \bar{\Psi}_p \gamma_5 \Psi_n \Phi
+  \sqrt{2}  \bar{\Psi}_n \gamma_5 \Psi_p \Phi^*
+  (  \bar{\Psi}_p \gamma_5 \Psi_p - \bar{\Psi}_n \gamma_5 \Psi_n ) \phi
\right] \nonumber\\
&-&  \frac{\beta_2 f\sqrt{2}}{m} \left[ \bar{\Psi}_p \gamma_5 
\gamma_\mu \left[ (\partial^\mu +ie A^\mu) \Phi \right] \Psi_n 
+\sqrt{2} \bar{\Psi}_n \gamma_5 
\gamma_\mu \left[ (\partial^\mu -ie A^\mu) \Phi^* \right] \Psi_p \right]  \nonumber \\
&-& \frac{\beta_2 f}{m} \left[ 
\bar{\Psi}_p \gamma_5 \gamma_\mu (\partial^\mu \phi) \Psi_p -
\bar{\Psi}_n \gamma_5 \gamma_\mu (\partial^\mu \phi) \Psi_n \right]\,,
\end{eqnarray}
with $\beta_1 + \beta_2 = 1$ . The generalized sources simply are 
\begin{eqnarray}
F_A^{M} &=& \beta_1 F + \beta_2 F_A^{PV}  \,
 , \nonumber \\
\tilde{F}_A^{M} &=& \beta_1 \tilde{F} + \beta_2\tilde{F}_A^{PV}  \,
 , \nonumber \\
F_0^M &=& \beta_1 F_0 + \beta_2 F_0^{PV} \, .
\end{eqnarray}
Herewith the results easily follow: just replace the earlier
generalized sources by the mixed ones. In particular, one
obtains for the effective Lagrangian
\begin{equation}
{\cal L}^{ef, ex}_{M} (x) = {\cal L}_1^{ex}(x) 
+ \Delta{\cal L}_0^{M}(x) 
+ \Delta{\cal L}_A^{M}(x) \, ,
\label{eq:LeffM}
\end{equation}
with
\begin{equation}
\Delta{\cal L}_0^{M}(x) = - \frac{1}{2}  \int d^4 y \, 
F_0^{M}(x) G_0(x-y) F_0^{M}(y) \, ,
\end{equation}
and
\begin{equation}
\Delta{\cal L}_A^{M}(x) =
- \int d^4 y \, \tilde{F}_A^{M}(x) G_A(x, y) F_A^{M}(y) \, .
\end{equation}
This concludes the formalism for external electromagnetic 
fields.

\section{Dynamical photons}
The effective actions derived in the previous section
describe nucleons in an external electromagnetic field.
For dynamical photons the formalism needs to be extended.
However, an exact extension is not possible yet and
therefore perturbation theory in the electromagnetic
coupling is applied. We present explicit results up
to (and including) $O(e^2)$.

The first step towards a full dynamical theory is trivial.
In the effective Lagrangians (cf. eqs. (\ref{eq:Leff}, \ref{eq:LeffPV},
\ref{eq:LeffM})  one merely replaces the `external field'
nucleon-photon Lagrangian, eq. (\ref{eq:nphoeff}), by the
original expression, eq. (\ref{eq:npho}). (Although the latter
Lagrangian contains a gauge fixing term, in the following
we will do a completely gauge invariant calculation.)
The problem arises due to the integration of the complex
pion field. Apart from the terms already given, the determinant
of the operator $O_A$ contributes to the effective action.
It represents a self-interaction term of the electromagnetic
field. In other words, the photon propagator is modified.
This is exactly what one physically expects: the charged
pions polarize the vacuum and also after integrating them
out the effect should be there.

Exploiting his proper-time
formalism, Schwinger \cite{Schw} explicitly calculated the analogous 
effective action due to the spin 1/2 field.
Moreover, he gave the result for the spin zero charged field. 
Nevertheless, we believe it to be useful to present 
our alternative derivation, which is -lacking the genius of Schwinger-
more straightforward but tedious and uses some modern techniques. 
Thus, in this respect, it may  render the present paper not only self-contained 
but the result also more accessible.   

Our calculation starts with the
famous trace-log formula for an operator $Q$,
%$\begin{equation}
$ \displaystyle
\det Q =\exp \left( \mbox{tr} \ln Q \right) \, .$
%\end{equation}
Application to the problem in question yields
the effective photon action 
%\begin{equation}
%\Delta S_A^{ef} \propto \mbox{tr} \ln(-G_A).
%\end{equation}
\begin{equation}
\Delta S_A^{ef} \propto \mbox{tr} \ln(-G_A)
- \mbox{tr} \ln(-G_0)
= \mbox{tr} \ln(G_0^{-1} G_A) \, ,
\end{equation}
where we have subtracted the corresponding expression for $G_0$,
in order to get rid of the first, trivial divergence.
We insert the perturbative result for $G_A$ derived in
Appendix A, and further expand in $e$ 
\begin{equation}
\Delta S_A^{ef} \propto  e \mbox{tr}\left({\cal D} G_0\right)
+ e^2 \mbox{tr}\left(\frac{1}{2} {\cal D} G_0 {\cal D} G_0 -A^2 G_0\right)
+O(e^3) \, .
\end{equation}
Since the appearing explicit expressions contain infinities
one has to regularize the theory. We choose the dimensional
regularization scheme which preserves gauge invariance.
The appearing momentum as well as space-time integrals are 
defined in $n$ dimensions and expanded around $n=4$. 
For dimensional reasons, one
also replaces $e$ by $ e M_0^{2-n/2}$, where $M_0$
is an arbitrary reference mass.  We refer to the textbooks,
e.g. \cite{Ry}, for more details and the original references.

Let us start with the first order term. The appearing
operator follows from eqs. (\ref{eq:fpro}) and (\ref{eq:Dper}),
%\begin{equation}
%{\cal D} G_0 (x, y) = i \left( \partial_x^\mu A_\mu \right)
%\int \frac{d^n k}{(2 \pi)^n} \frac{\exp ik(x-y)}{k^2-m^2+i\epsilon}
%- 2 A^\mu (x) 
%\int \frac{d^n k}{(2 \pi)^n} \frac{ k_\mu \exp ik(x-y)}{k^2-m^2+i\epsilon} \, .
%\end{equation}
and since $\mbox{tr}\, Q = \int d^n x \, Q(x, x)$ we get
\begin{equation}
\mbox{tr} \left({\cal D} G_0\right) = \int d^n x \, \left[
 \left(i\partial_x^\mu A_\mu \right)
\int \frac{d^n k}{(2 \pi)^n} \frac{1}{k^2-m^2+i\epsilon}
- 2 A^\mu (x) 
\int \frac{d^n k}{(2 \pi)^n} \frac{ k_\mu}{k^2-m^2+i\epsilon} \right] = 0 \, .
\end{equation}
The first term vanishes because $ \int d^n x \, 
 \left(i\partial_x^\mu A_\mu \right) =0 \, , $ the second one because
$ \int \frac{d^n k}{(2 \pi)^n} \frac{ k_\mu}{k^2-m^2+i\epsilon} =0 \, .$ 
Consequently,
% the first order term is identically zero:
%\begin{equation}
%\mbox{tr} \left({\cal D} G_0\right) = 0 \, .
%\end{equation}
no $O(e)$ term in the effective photon action
is generated. In other words, the vacuum is not polarized
in  first order.

The calculation of the second order term is more strenuous. 
Putting in the operators and integration by parts yields
\begin{eqnarray}
\mbox{tr}\left(\frac{1}{2} {\cal D} G_0 {\cal D} G_0 -A^2 G_0\right)
&=& \int d^n x \, \int d^n z \, \int \frac{d^n l}{(2 \pi)^n}
\, \int \frac{d^n q}{(2 \pi)^n} A_\mu(x) A_\nu(z)
\frac{\exp il(x-z)}{l^2-m^2+i\epsilon}
\frac{\exp iq(z-x)}{q^2-m^2+i\epsilon} 
\nonumber \\
&\cdot&\left[ 2 l^\mu q^\nu -  l^\mu (q-l)^\nu - q^\nu (l-q)^\mu
+\frac{1}{2}(l-q)^\mu (q-l)^\nu \right] \nonumber \\
&-& \int d^n x \, A^2(x) \int \frac{d^n k}{(2 \pi)^n}
\frac{1}{k^2-m^2-i\epsilon} \, . 
\label{eq:moin}
\end{eqnarray}
%We use the Fourier-representation of the vector field,
%$\displaystyle  A_\mu (x) =
 %\int \frac{d^n k'}{(2 \pi)^n} A_\mu(k') \exp ik'x \,$,
%in order to do the integrals over space-time 
%\begin{eqnarray}
%\mbox{tr}\left(...\right)
%%\mbox{tr}\left(\frac{1}{2} {\cal D} G_0 {\cal D} G_0 -A^2 G_0\right)
%&=&  \int \frac{d^n l}{(2 \pi)^n}
%\, \int \frac{d^n q}{(2 \pi)^n} A_\mu(q-l) A_\nu(l-q)
%\frac{1}{l^2-m^2+i\epsilon}
%\frac{1}{q^2-m^2+i\epsilon} 
%\nonumber \\
%&\cdot&\left[ 2 l^\mu q^\nu +  l^\mu (l-q)^\nu + q^\nu (q-l)^\mu
%+\frac{1}{2}(l-q)^\mu (q-l)^\nu \right] \nonumber \\
%&-& \int \frac{d^n p}{(2 \pi)^n} \, A_\mu (-p) A^\mu (p)
%\int \frac{d^n k}{(2 \pi)^n} \frac{1}{k^2-m^2-i\epsilon} \, . 
%\end{eqnarray}
%By introducing the new variables $ p = l-q, \; 2k =l+q\, ,$ and 
%combining the denominators by means of Feynman's trick, we obtain
%\begin{eqnarray}
%\mbox{tr}\left(...\right)
%&=& 
 %2 \int \frac{d^n p}{(2 \pi)^n} \, A_\mu(-p) A_\nu(p)
%\int_0^1 dx \,\int \frac{d^n k}{(2 \pi)^n}
%\frac{ k^\mu k^\nu}{(k^2+2(x-\frac{1}{2})k p +\frac{1}{4}p^2-m^2+i\epsilon)^2}
%\nonumber \\
%&-& \int \frac{d^n p}{(2 \pi)^n} \, A_\mu (-p) A_\nu (p) g^{\mu \nu}
%\int \frac{d^n k}{(2 \pi)^n} \frac{1}{k^2-m^2-i\epsilon}  
%\,.
%\label{eq:moin}
%\end{eqnarray}
%At this point we recognize the integrals of the one-loop
%vacuum polarization in scalar electrodynamics (see e.g. \cite{IZ}).
%The second term corresponds to the ``tad-pole'' diagram.

These integations can be done and the results expanded
around $4-n$. We refer to Appendix B for the technical details
and immediately proceed to the  result, the effective regularized action
%\begin{equation}
$S_A^{ef} = S_A^0 + \Delta S_A^{ef},$
%\end{equation}
where (in momentum space, omitting $O(4-n)$ and $O(e^3)$)
\begin{eqnarray}
\Delta S_A^{ef} &=& \frac{e^2 \pi^2}{3(2 \pi)^4} \int \frac{d^n p}{(2 \pi)^n}
A_\mu (-p) A_\nu(p) \left[ p^\mu p^\nu - p^2 g^{\mu \nu} \right] \nonumber \\
&\cdot& \left[\frac{1}{4-n} -\frac{1}{2}\gamma_E-
\frac{3}{2} \int_0^1 du \, (1-2u)^2
\ln\left(\frac{m^2-u(1-u)p^2 -i\epsilon}{4\pi M_0^2}\right) \right] \, ,
\end{eqnarray}
with Euler's constant $\gamma_E= 0.577..\; .$
Recall the free action
\begin{equation}
S_A^0 = -\frac{1}{4} \int d^n x \, F_{\mu \nu}(x) F^{\mu \nu}(x)
= -\frac{1}{4} \int \frac{d^n k}{(2 \pi)^n} F_{\mu \nu}(-k) F^{\mu \nu}(k) \, ,
\end{equation}
with $F_{\mu \nu}(k) =ik_\mu A_\nu (k)-i k_\nu A_\mu (k)\,.$ 
Verifying
%\begin{equation}
$F_{\mu \nu}(-k) F^{\mu \nu}(k)  = 2 A_\mu (-k) A_\nu (k) 
\left[ k^2 g^{\mu \nu} - k^\mu k^\nu \right] \, ,$
%\end{equation}
explicitly demonstrates the local gauge invariance of $\Delta S_A^{ef}$,
including the (for $n=4$) divergent terms. It also enables us
to rewrite the effective action as
\begin{eqnarray}
S_A^{ef} &=& -\frac{1}{4} \int \frac{d^n k}{(2 \pi)^n}
F_{\mu \nu}(-k) F^{\mu \nu}(k) \nonumber \\
&\cdot& \left(
1+ \frac{2 e^2 \pi^2}{3(2 \pi)^4} \left[
\frac{1}{4-n} -\frac{1}{2}\gamma_E-
\frac{3}{2} \int_0^1 du \, (1-2u)^2
\ln\left(\frac{m^2-u(1-u)k^2 -i\epsilon}{4\pi M_0^2}\right) \right] \right) \,.
\end{eqnarray}

The latter form is also suited to renormalize the theory. It shows
that the term we need to substract, {\it i.e.}, the counterterm,
indeed has the same structure as the original action.
As is usual in electrodynamics we perform `on-shell' renormalization.
Thus we choose the subtraction point to coincide with the
physical photon mass, $k^2 = 0$, and (apart from cancelling
the divergence) demand that the pole of the renormalized
propagator remains at $k^2=0$ with residue $1$. This uniquely
fixes the subtraction to be
\begin{equation}
\delta S_A = -\frac{1}{4} \int \frac{d^n k}{(2 \pi)^n}
F_{\mu \nu}(-k) F^{\mu \nu}(k) 
 \frac{2 e^2 \pi^2}{3(2 \pi)^4} \left[
\frac{1}{4-n} -\frac{1}{2}\gamma_E
+\frac{1}{2}\ln\left(\frac{m^2}{4\pi M_0^2}\right) \right] \,.
\end{equation}
In this way we get for the effective renormalized action
\begin{eqnarray}
S_A^{ef} &=& -\frac{1}{4} \int \frac{d^4 k}{(2 \pi)^4}
F_{\mu \nu}(-k) F^{\mu \nu}(k) \nonumber \\
&\cdot& \left( 1 -
 \frac{e^2 \pi^2}{(2 \pi)^4} \left[ \int_0^1 du \,
(1-2u)^2 \ln\left(\frac{m^2-u(1-u)k^2-i\epsilon}{4\pi M_0^2}\right)
-\frac{1}{3}\ln\left(\frac{m^2}{4\pi M_0^2}\right) \right] \right) \,.
\end{eqnarray}
After integration by parts, hereby cancelling the reference mass $M_0$,
%\begin{equation}
 %\int_0^1 du \, (1-2u)^2
%\ln\left(\frac{m^2-u(1-u)k^2-i\epsilon}{4\pi M_0^2}\right) =
%\frac{1}{3}\ln\left(\frac{m^2}{4 \pi M_0^2}\right) - \frac{1}{6} k^2
 %\int_0^1 du \, \frac{(1-2u)^4}{m^2-u(1-u)k^2-i\epsilon} \, ,
%\end{equation}
%shows that the reference mass $M_0$ cancels. 
and by changing the integration
variable to $v=2u-1\, ,$ we finally obtain the finite and gauge invariant
result \cite{Schw}
\begin{equation}
S_A^{ef} = -\frac{1}{4} \int \frac{d^4 k}{(2 \pi)^4}
F_{\mu \nu}(-k) F^{\mu \nu}(k) \,
 \left[ 1 + \frac{k^2}{6 m^2}
 \frac{e^2}{(4 \pi)^2}  \int_0^1 dv \,
\frac{v^4}{1-\frac{k^2}{4 m^2}(1-v^2)-i\epsilon} \right] \,.
\end{equation}
We kept the $i \epsilon$ prescription because the denominator in
the equation above can
vanish for $k^2 \ge 4 m^2$. This corresponds to  pion
pair production by the electromagnetic field.

\section{Local gauge invariance}

The results of the previous section, in particular the effective
photon action, is manifestly gauge invariant. The local
gauge invariance of the nonlocal Lagrangians derived in
section (3) is not that obvious. Therefore, we
want to verify this; it reduces to explicitly check
the invariance of the additional terms 
$\Delta{\cal L}_0, \Delta{\cal L}_A,
\Delta{\cal L}_0^{PV}, \Delta{\cal L}_A^{PV},
\Delta{\cal L}_0^{M}, \Delta{\cal L}_A^{M}$
without external sources. 

It is easily seen that the generalized sources (with the
external ones put to zero) have the following 
transformation properties under local gauge transformations:
\begin{eqnarray}
\left( F, F_A^{PV}, F_{A}^M \right)(x)_{J=0} 
&\rightarrow&  \exp\left(-ie\chi (x)\right)
\left( F, F_A^{PV}, F_{A}^M \right)(x)_{J^=0} \, ,  \nonumber \\
\left( \tilde{F}, \tilde{F}_A^{PV}, \tilde{F}_{A}^M \right)(x)_{J^*=0} 
&\rightarrow&  \exp\left(ie\chi (x)\right)
\left( \tilde{F}, \tilde{F}_A^{PV}, \tilde{F}_{A}^M \right)(x)_{J^*=0}
 \, ,\nonumber \\
\left( F_0, F_0^{PV}, F_{0}^M \right)(x)_{J_0=0} 
&\rightarrow&  
\left( F_0, F_0^{PV}, F_{0}^M \right)(x)_{J_0=0} \, . 
\end{eqnarray}
The third relation immediately guarantuees the invariance of
$\Delta{\cal L}_0, \Delta{\cal L}_0^{PV}, $ and
$\Delta{\cal L}_0^{M},$ corresponding to neutral pions.
The first two transformations, however, require
\begin{equation}
G_A(x, y) \rightarrow G_{A+\partial \chi} = \exp\left(-ie\chi(x)\right)
G_A(x,y) \exp\left(ie\chi(y)\right) \, ,
\label{eq:GC}
\end{equation}
in order  that  
$\Delta{\cal L}_A, \Delta{\cal L}_A^{PV}, $ and $\Delta{\cal L}_A^{M}$
are  locally gauge invariant.

We first verify this relation nonperturbatively by only using the definition
of $G_A$, eq. (\ref{eq:OA}). After a gauge transformation one has
\begin{equation}
{\cal O}_{A+\partial \chi} (x) G_{A+\partial \chi} (x,y) = -\delta^4(x-y) \, .
\end{equation}
Multiplication with the local phase factors gives
\begin{equation}
\exp\left(ie\chi(x)\right) \exp\left(-ie\chi(y)\right)
{\cal O}_{A+\partial \chi} (x) G_{A+\partial \chi} (x,y) = 
-\delta^4(x-y) \, .
\end{equation}
With some algebra one can show
\begin{equation}
{\cal O}_A \left[ \exp\left(ie\chi(x)\right) G_{A+\partial \chi}
 \exp\left(-ie\chi(y)\right) \right] =
\exp\left(ie\chi(x)\right) \exp\left(-ie\chi(y)\right)
{\cal O}_{A+\partial \chi} (x) G_{A+\partial \chi}(x,y) \, .
\end{equation}
Combining the latter two equations yields
\begin{equation}
\left[ \exp\left(ie\chi(x)\right) G_{A+\partial \chi}
 \exp\left(-ie\chi(y)\right) \right] = G_A(x,y) \, ,
\end{equation}
where also uniqueness of the
Greens function has been used. This completes the proof.    

However, recall that we only can provide a perturbative construction
of the operator $G_A$. Do we indeed satisfy local gauge invariance
order by order in $e\,$? In order to answer this question one needs to
expand the gauge condition, eq. (\ref{eq:GC}), using the expansion
for $G_A$ (see Appendix A).
For the gauge transformed $G_n$
we introduce the notation
\begin{equation}
G_n \rightarrow G_n + \delta G_n \, .
\end{equation}
Local gauge invariance to
zeroth order is trivial; in first order we need to show that
\begin{equation}
i\left[\chi(x)-\chi(y)\right] G_0(x-y) + \delta G_1(x, y) = 0 \, .
\label{eq:GC1}
\end{equation}
Explicitly we have for $\delta G_1$,
\begin{equation}
\delta G_1 (x,y) = i \int d^4 z \, G_0(x-z)\left[
2 \left(\partial^z_\mu \chi (z) \right) \partial_z^\mu
+\left(\partial^\mu_z \partial_\mu^z \chi(z) \right) \right] G_0(z-y) \, .
\end{equation}
After integration by parts and putting in the definition of the free propagator,
eq. (\ref{eq:fprod}), we readily obtain eq. (\ref{eq:GC1}).
The second order term is more involved. The gauge condition reads 
\begin{equation}
\frac{1}{2} \left[ \chi(x) - \chi(y) \right]^2 G_0(x, y) + 
i\left[\chi(x)-\chi(y)\right] G_1(x,y) + \delta G_2(x, y) = 0 \, .
\label{eq:GC2}
\end{equation}
Since $G_2= G_0 {\cal D} G_1 - G_0 A^2 G_0 $ (see Appendix A),
we get
\begin{eqnarray}
\delta G_2(x, y) &=& \int d^4 z \, G_0(x-z) \big\{ -i {\cal D}(z) 
\left(\chi(z)-\chi(y)\right)G_0(z-y)
\nonumber \\ &+& i
 \left(2(\partial_\mu^z\chi(z))\partial^\mu_z
+  \left(\partial_\mu^z \partial^\mu_z \chi(z)\right)\right)G_1(z, y)
\nonumber \\
&+&\big[ \left(2(\partial_\mu^z\chi(z))\partial_z^\mu
+\left(\partial^z_\mu \partial_z^\mu \chi(z)\right)\right)\left(\chi(z)-\chi(y)\right)
\nonumber \\ &-&
 2(\partial_\mu^z \chi(z))A^\mu(z) +
\left(\partial_\mu^z \chi(z)\right)\left(\partial^\mu_z \chi(z)\right)\big]G_0(z-y) 
\big\} \,.
\end{eqnarray}
Now it is straightforward but somewhat tedious to verify eq. (\ref{eq:GC2}).
Most conveniently one separates linear and quadratic terms in $\chi$. 
Furthermore, only integration by parts and the defining differential
equations for $G_0$ and $G_1$ are needed. If some reader feels the
inspiration, he/she is invited to show the local gauge invariance
of the higher order terms. Note, however, that we gave
a nonperturbative proof; in other words, these explicit verifications
only serve as a check of the perturbative  calculations.

We conclude this section by re-emphasizing that we have achieved to
eliminate the pion degrees of freedom while maintaining 
local gauge invariance. Moreover, the induced effective strong and
electromagnetic interactions are consistent and (given the
original action) unambiguous. Furthermore, the results obtained
so far are formally exact, in particular nonperturbative in
the strong coupling constant. Because we have obtained a 
nonlocal field theory with four-fermion interactions, its
applicability as a relativistic quantum field theory may be limited.

\section{Nonrelativistic reduction}

\subsection{Propagators}
In the nonrelativistic limit ($c \rightarrow \infty$) the 
differential operators ${\cal O}$ and ${\cal O}_A$ reduce to
\begin{eqnarray}
{\cal O} \rightarrow {\cal O}^{N} &=& m^2 - \Delta  \, , \nonumber \\
{\cal O}_A \rightarrow {\cal O}^{N}_A &=& {\cal O}^{N} + e {\cal D}^{N}
-e^2 \vec{A}^2 \, ,
\end{eqnarray}
where $O(1/c^2)$ terms have been neglected and
\begin{equation}
{\cal D}^{N} = 2i A_k \partial^k + i (\partial^k A_k) \, .
\end{equation}
Note that ${\cal O}_A^N$ is time dependent but does not contain
time derivatives. Concomitantly, it does not depend on the scalar potential
$A_0$; the indices $A$ below therefore
refer to the vector potential $\vec A$ only.
The inverse operators satisfy
\begin{eqnarray}
{\cal O}^{N} G_0^{N}(x, y) &=& - \delta^4(x-y)  \, , \nonumber \\
{\cal O}^{N}_A G_A^{N}(x, y) &=& - \delta^4(x-y) \, .
\end{eqnarray}
We define nonrelativistic propagators by separating the
delta function in time:
\begin{eqnarray}
 G_0^{N}(x, y) &=& - \delta(x^0-y^0)\, g_0(\vec{x}, \vec{y})  \, , \nonumber \\
 G_A^{N}(x, y) &=& - \delta(x^0-y^0)\, g_A(\vec{x}, \vec{y}, t)\, ,
\end{eqnarray}
which immediately yields
\begin{eqnarray}
{\cal O}^{N} g_0(\vec{x}, \vec{y}) &=& \delta^3(\vec x-\vec y) \, ,\nonumber \\
{\cal O}^{N}_A g_A(\vec x, \vec y, t) &=& \delta^3(\vec x-\vec y) \, .
\label{eq:GNA}
\end{eqnarray}
Here we `divided out' $\delta(x^0-y^0)$, which apparently
renders this procedure
to be nonunique since the nonrelativistic propagators can
be multiplied by a function $f(x^0, y^0)$ with $f(t, t) =1$.
However, these ambiguities will disappear
via the integration over $y^0$.

The solution for the free static propagator is well-known,
\begin{equation}
g_0(\vec x, \vec y) = \int \frac{d^3 k}{(2 \pi)^3} \, \frac{\exp (-i \vec k
(\vec x - \vec y))}{\vec k ^2 +m^2} =
\frac{\exp\left(- m |\vec x - \vec y| \right)}{4\pi |\vec x -\vec y|}  \, .
\end{equation}
For arbitrary electromagnetic fields we again can construct $g_A$
only perturbatively in $e$. The actual construction is completely
analogous to the relativistic case (see Appendix A) and starts
with the expansion
\begin{equation}
g_A (\vec x, \vec y, t) = \sum_{n=0}^{\infty} e^n
g_n(\vec x, \vec y, t) \, .
\end{equation}
We readily obtain
\begin{equation}
g_1 = -g_0 {\cal D}^N g_0 \, .
\end{equation}
Here and in the following the integrations are three-dimensional;
thus explicitly $g_1$ reads
\begin{equation}
g_1(\vec x, \vec y, t)  = - \int d^3 z \, g_0(\vec x, \vec z)
 {\cal D}^N(\vec z, t) g_0(\vec z, \vec y) \, .
\end{equation}
For $n \ge 2$ we get the recursion relation
\begin{equation}
g_n = -g_0 \left({\cal D}^N g_{n-1} - \vec A ^2 g_{n-2} \right) \, ,
\end{equation}
allowing for a calculation of $g_A$ to arbitrary order in $e$.

Let us address local gauge transformations. The transformation
property of the relativistic progator $G_A$, given in eq. (\ref{eq:GC}),
actually suggests
\begin{equation}
g_A(\vec x, \vec y, t) \rightarrow g_{A+\partial \chi} =
\exp\left(-ie\chi(\vec x, t)\right)
g_A(\vec x, \vec y, t) \exp\left(ie\chi(\vec y, t)\right) \, .
\end{equation}
Analogous to the relativistic case one can verify this
relation either exactly, by using the defining differential
equation (\ref{eq:GNA}), or perturbatively by means of the
explicit construction given above.
\subsection{Fermion fields}
The nonrelativistic approximation of the nucleon fields
is obtained in the standard way, see e.g. {\cite{BD}}.
One considers the Dirac equation in an electromagnetic field
and writes the fields as
\begin{equation}
\Psi_{p, n} = \left( \begin{array}{c} \tilde{\phi}_{p, n}(x) \\
\tilde{\chi}_{p, n}(x) \end{array}\right) = e^{-iMt}
\left( \begin{array}{c} \phi_{p, n} (x) \\
\chi_{p, n} (x) \end{array}\right) \, ,
\end{equation}
where $\phi$ and $\chi$ are two-component spinors. Then one
approximately has
\begin{eqnarray}
\chi_p(x) &\simeq& \frac{\vec{\sigma} \cdot \vec{\pi}}{2M} \phi_p(x) \, , 
\hspace{2.0cm}
\chi^{\dagger}_p(x) \simeq - \frac{ \underline{\vec{\pi}}}{2M}
\phi^{\dagger}_p(x) \cdot \vec{\sigma} \, , 
\nonumber \\
\chi_n(x) &\simeq& \frac{\vec{\sigma} \cdot \vec{p}}{2M} \phi_n(x) \, , 
\hspace{2.0cm}
\chi^{\dagger}_n(x) \simeq - \frac{ \vec{p}}{2M}
\phi^{\dagger}_n(x) \cdot \vec{\sigma} \, ,
\end{eqnarray}
with $\vec{\pi} = \vec{p}-e\vec{A}(x) , \underline{\vec{\pi}}=\vec{p}
+e\vec{A}(x)\; \mbox{and } \vec{p} = -i \nabla \, .$ In case there are
time derivatives, factors $M$ appear; for instance:
\begin{eqnarray}
\left( i\partial_t - e A_0\right) \tilde{\phi}_p &=&
M e^{-iMt} \phi_p + e^{-iMt} \left( i\partial_t -e A_0\right) \phi_p \, ,
\nonumber \\
\left( i\partial_t - e A_0\right) \tilde{\chi}_p &=&
-M e^{-iMt} \chi_p + e^{-iMt} \left(\vec{\sigma} \cdot \vec{\pi}\right)
\phi_p \, .
\end{eqnarray}
Analogous expressions for neutron and/or hermitian conjugate
fields can readily be derived.

At this point the nonrelativistic reduction of the effective Lagrangians is 
straightforward.  Some remarks, however, may be appropriate. Since
the Dirac equation including electromagnetic field has been used
above, the reduction is locally gauge invariant. On the other hand, the
procedure obviously does not take into account the strong interaction terms 
in the equation of motion. Thus it is implicitly assumed that
the strong interaction energy is small compared to the mass and therefore
can be neglected in this order.
Although this is common practice, it renders
this nonrelativistic approximation somewhat uncontrolled in this case. 
\subsection{Nonrelativistic action}
With the ingredients derived in the previous sections we obtain as
nonrelativistic limit of the nucleon-photon
Lagrangian ${\cal L}_1$, eq. (\ref{eq:npho}),
\begin{eqnarray}
%{\cal L}_1 &\simeq& \phi_p^{\dagger}\left(i\partial_0 - eA_0\right) \phi_p
{\cal L}_1^{NR} &=& \phi_p^{\dagger}\left(i\partial_0 - eA_0\right) \phi_p
-\frac{1}{2M} \phi_p^{\dagger}  (\vec{\sigma}\cdot \vec{\pi})
 (\vec{\sigma}\cdot \vec{\pi})\phi_p 
 + \phi_n^{\dagger}i\partial_0  \phi_n
-\frac{1}{2M} \phi_n^{\dagger} (\vec{\sigma} \cdot \vec{p})
 (\vec{\sigma} \cdot \vec{p})\phi_n - \frac{1}{4} F_{\mu \nu}F^{\mu \nu} 
\nonumber \\
 &=& \phi_p^{\dagger}\left(i\partial_0 - eA_0\right) \phi_p
-\frac{1}{2M} \phi_p^{\dagger} (\vec{\pi}^2
 -e \vec{\sigma} \cdot \vec{B})\phi_p 
 + \phi_n^{\dagger}i\partial_0  \phi_n
-\frac{1}{2M} \phi_n^{\dagger} \vec{p}^2 \phi_n 
+\frac{1}{2}(\vec{E}^2-\vec{B}^2) \, .
\end{eqnarray}
Here and in the following we neglect $O(1/c^2)$ contributions.

We proceed by approximating the generalized sources; for
pseudoscalar coupling and zero external sources  we easily get
\begin{eqnarray}
F &\simeq& -ig\sqrt{2}\left(\phi_n^{\dagger} \frac{\vec{\sigma} \cdot \vec{\pi}}{2M}
\phi_p + \frac{\vec{p}}{2M}\phi_n^{\dagger} \cdot \vec{\sigma} \phi_p\right)
\, , \nonumber \\
\tilde{F} &\simeq& -ig\sqrt{2}\left(\phi_p^{\dagger} \frac{\vec{\sigma}
\cdot \vec{p}}{2M}
\phi_n + \frac{\underline{\vec{\pi}}}{2M}\phi_p^{\dagger} \cdot
\vec{\sigma} \phi_n\right)
\, , \nonumber \\
F_0 &\simeq& -ig\left(\phi_p^{\dagger} \frac{\vec{\sigma} \cdot \vec{p}}{2M}
\phi_p + \frac{\vec{p}}{2M}\phi_p^{\dagger} \cdot \vec{\sigma} \phi_p
-\phi_n^{\dagger} \frac{\vec{\sigma} \cdot \vec{p}}{2M}
\phi_n - \frac{\vec{p}}{2M}\phi_n^{\dagger} \cdot \vec{\sigma} \phi_n\right)\, .
\end{eqnarray}
Note that these nonrelativistic sources depend on the gauge field. This
is due to the (gauge-invariant) elimination of $\chi_p$.

The analogous calculation for pseudovector coupling is more
involved because these sources contain derivatives. Nevertheless,
the final result is simple. In fact, using the relation
between the coupling constants $g$ and $f$, {\it i.e.},
$\frac{f}{m}=\frac{g}{2M}$ 
yields
\begin{eqnarray}
F_A^{PV} &=& F  +O(1/c^2)
\, , \nonumber \\
\tilde{F}_A^{PV} &=& \tilde{F} +O(1/c^2)
\, , \nonumber \\
F_0^{PV} &=& F_0 +O(1/c^2) \, .
\end{eqnarray}
Therefore, also mixed coupling produces the same result.

As a consequence, irrespective of choosen strong coupling (pseudoscalar,
pseudovector or mixed),
we find an unique lowest order result for the nonrelativistic
reduction of the effective Lagrangians:
\begin{equation}
{\cal L}^{NR}(x) = {\cal L}_1^{NR}(x) +\frac{1}{2} \int d^3 y \, F_0(x)
g_0(\vec{x} - \vec{y}) F_0(y)
+\int d^3 y \, \tilde{F}(x) g_A(\vec{x}, \vec{y}, t) F(y) \, ,
\end{equation}
with $x^0=y^0=t$.
We can include the effective photon action derived
in section (3), 
\begin{equation}
S^{NR}= S_A^{eff} + \int d^4 x \, {\cal L}^{NR}(x) + \frac{1}{4}F_{\mu \nu}
F^{\mu \nu} \, , 
\end{equation}
where we need to subtract the free photon term 
from ${\cal L}$ because it is
already contained in $S_A^{eff}$.  This action is locally
gauge invariant. We may add the nonrelativistic form of the
gauge invariant Pauli
terms in order to account for the anomalous magnetic moments of
the nucleons. Furthermore, one may reintroduce
the external sources and/or the gauge fixing term. 
In any case, this final result represents a nonrelativistic field theory
for interacting nucleons, which also interact with the
electromagnetic field. 

%HERE

\section{Summary and outlook}
Using functional integrations,
we have shown a way to eliminate pion degrees
of freedom from a relativistic quantum field theory describing
nucleons, pions and dynamical photons. 
This elimination can be done exactly, {\it i.e.}, nonperturbatively in the
strong coupling constant, as long as 
the pion-nucleon interactions are linear in the pion fields. 
Indeed we have considered pseudoscalar, pseudovector and
`mixed' pion-nucleon interactions. The induced photon-photon
interaction is only calculated up to order $e^2$ and
agrees with the result of Schwinger \cite{Schw}. Another appearing
operator, accounting for charged pion propagation in an electromagnetic
field, is constructed perturbatively in $e$.

After having integrated out the pions, we have obtained a nonlocal
relativistic quantum field theory which is nevertheless
locally gauge invariant. Effective, mutually consistent,
strong and electromagnetic interactions have appeared. Furthermore,
the nucleons have acquired structure due to the pion cloud.
At this point, it is not clear whether this intermediate result is suited
for detailed practical calculations. For instance,
the nonlocalities may prevent applications beyond the tree level.
Further studies, adressing regularization and renormalization
of the nonlocal action, are therefore desirable.

Here we have subsequently made a nonrelativistic reduction in
order to arrive at a nonrelativistic quantum field theory
for nucleons interacting with the (dynamical) electromagnetic field.
In this nonrelativistic limit, the resulting action is unique
irrespective of choosing pseudoscalar, pseudovector or mixed pion-nucleon
coupling in the original action. Potentials, electromagnetic
structure as well as 
exchange currents are consistently generated. The action is invariant
under $U(1)$ local gauge transformations.
This symmetry  has several consequences. First,
no ambiguities concerning exchange contributions in order
to restore gauge invariance arise. Secondly, the nonrelativistic
Ward-Takahashi \cite{Nabo} holds. Finally, as also can be seen
explicitly, Siegert's hypothesis is satisfied. 

The resulting nonrelativistic field theory can be used to
derive the Hamiltonian of the $N$-nucleon system, including
the electromagnetic interactions. For $N=1$ electromagnetic
structure, {\it i.e.}, form factors, will be isolated whereas for
$N=2$ the effective interaction and two-body currents will
explicitly appear. In particular, it will be interesting to compare
to other, existing, results for these exchange contributions.
The formalism presented here may also provide a systematic way
to derive relativistic corrections.
Moreover, it can be applied to the analogous scalar
and vector meson exchange models.
We hope to address these issues in the nearby future.

\section*{Acknowledgments}
The author acknowledges the suggestion of F. Lenz to apply
the functional formalism to these problems. He thanks K. Bugaev, P. U. Sauer
and T. Wilbois for useful discussions and/or critical
readings of the manuscript.

\setcounter{equation}{0}
\renewcommand{\theequation}{A.\arabic{equation}}
\section*{Appendix A: Pion propagation in an electromagnetic
field}
\appendix
We perturbatively construct the
inverse operator of ${\cal O}_A$, {\it i.e.}, $G_A (x, y)$
(cf. eq. (\ref{eq:OA})). Since the result is inductive,
this method yields $G_A$ to any arbitrary order in the
electromagnetic coupling constant $e$.

Let us start by assuming the following expansion 
\begin{equation}
G_A (x,y) = \sum_{n=0}^{\infty} e^n G_n(x,y) \, .
\end{equation}
The $G_n$ are to be determined from the differential
equation (cf. eqs. (\ref{eq:OA}, \ref{eq:Oper}, \ref{eq:Dper}))
\begin{equation}
\left[ {\cal O} + e {\cal D} - e^2 A^2 \right] \sum_{n=0}^{\infty} e^n G_n(x,y)
= - \delta^4 (x-y) \, .
\end{equation}
The zeroth order solution is indeed nothing else than the free propagator,
\begin{equation}
G_0(x,y) = G_0(x-y) \, ,
\end{equation}
which has been explicitly given in eq. (\ref{eq:fpro}).
To first order we find
\begin{equation}
G_1 = G_0 {\cal D} G_0 \, ,
\end{equation}
which explicitly means
\begin{equation}
G_1(x, y)  = \int d^4 z \,  G_0(x-z) {\cal D}(z) G_0(z-y) \, .
\end{equation}
The next term follows as
\begin{equation}
G_2 = G_0 {\cal D} G_1 -G_0 A^2 G_0 = G_0{\cal D} G_0 {\cal D} G_0
-G_0 A^2 G_0  \, .
\end{equation}
Finally, one can easily verify the following recursion relation ($n \ge 2$) 
\begin{equation}
G_n = G_0 {\cal D} G_{n-1} -G_0 A^2 G_{n-2} \, .
\end{equation} This completes the construction of the propagator $G_A$.

\setcounter{equation}{0}
\renewcommand{\theequation}{B.\arabic{equation}}
\section*{Appendix B: Integrations}
First we use the Fourier-representation of the vector field,
$\displaystyle  A_\mu (x) =
 \int \frac{d^n k'}{(2 \pi)^n} A_\mu(k') \exp ik'x \,$,
in order to do the integrals over space-time in eq. (\ref{eq:moin})
\begin{eqnarray}
\mbox{tr}\left(...\right)
&=&  \int \frac{d^n l}{(2 \pi)^n}
\, \int \frac{d^n q}{(2 \pi)^n} A_\mu(q-l) A_\nu(l-q)
\frac{1}{l^2-m^2+i\epsilon}
\frac{1}{q^2-m^2+i\epsilon} 
\nonumber \\
&\cdot&\left[ 2 l^\mu q^\nu +  l^\mu (l-q)^\nu + q^\nu (q-l)^\mu
+\frac{1}{2}(l-q)^\mu (q-l)^\nu \right] \nonumber \\
&-& \int \frac{d^n p}{(2 \pi)^n} \, A_\mu (-p) A^\mu (p)
\int \frac{d^n k}{(2 \pi)^n} \frac{1}{k^2-m^2-i\epsilon} \, . 
\end{eqnarray}
Introducing the new variables $ p = l-q, \; 2k =l+q\, ,$ and 
combining the denominators by means of Feynman's trick, yield
\begin{eqnarray}
\mbox{tr}\left(...\right)
&=& 
 2 \int \frac{d^n p}{(2 \pi)^n} \, A_\mu(-p) A_\nu(p)
\int_0^1 dx \,\int \frac{d^n k}{(2 \pi)^n}
\frac{ k^\mu k^\nu}{(k^2+2(x-\frac{1}{2})k p +\frac{1}{4}p^2-m^2+i\epsilon)^2}
\nonumber \\
&-& \int \frac{d^n p}{(2 \pi)^n} \, A_\mu (-p) A_\nu (p) g^{\mu \nu}
\int \frac{d^n k}{(2 \pi)^n} \frac{1}{k^2-m^2-i\epsilon}  
\,.
\end{eqnarray}
These momentum integrals 
indeed can be evaluated in $n$ dimensions \cite{Ry}. We obtain
\begin{equation}
 \int \frac{d^n k}{(2 \pi)^n}
\frac{ k^\mu k^\nu}{(k^2+2(x-\frac{1}{2})k p +\frac{1}{4}p^2-m^2+i\epsilon)^2}=
\frac{i\pi ^{n/2}}{(2 \pi)^n} \frac{\Gamma(2-\frac{n}{2})}{\Gamma(2)}
f^{\frac{n}{2}-2}
\left[ (x-\frac{1}{2})^2 p^\mu p^\nu + 
\frac{f}{n-2} g^{\mu \nu}
 \right]\, , 
\end{equation}
with
\begin{equation}
f = f(p, x)= p^2(x-\frac{1}{2})^2 -
\frac{1}{4}p^2 +m^2 -i\epsilon =m^2 -x(1-x)p^2 -i \epsilon \, .
\end{equation}
At this point it is convenient to include the factor
$M_0^{4-n}$, which stems from the coupling constant, and the factor 2. 
Moreover, we also consider the $x$-integration; we get
\begin{eqnarray}
2 M_0^{4-n} \int_0^1 dx\, \int \frac{d^n k}{(2 \pi)^n}
\frac{ k^\mu k^\nu}{(k^2+2(x-\frac{1}{2})k p +\frac{1}{4}p^2-m^2+i\epsilon)^2}=
\nonumber \\
\frac{i\pi^2}{(2\pi)^4} \int_0^1 dx\,
\left(\frac{f}{4\pi M_0^2}\right)^{\frac{n-4}{2}}
\left[ \frac{1}{2}\Gamma (2-\frac{n}{2}) (1-2x)^2 p^\mu p^\nu
- \Gamma (1-\frac{n}{2}) f g^{\mu \nu}\right] \, .
\end{eqnarray}
The second momentum integral gives
\begin{equation}
g^{\mu \nu} M_0^{4-n} \int \frac{d^n k}{(2 \pi)^n}
\frac{1}{k^2-m^2+i\epsilon }= g^{\mu \nu} 
\frac{i\pi^2}{(2 \pi)^4} \Gamma (1-\frac{n}{2}) m^2
\left( \frac{m^2}{4 \pi M_0^2} \right)^{\frac{n-4}{2}} \, .
\end{equation}
As is clear from power counting, both integrals diverge
for $n \rightarrow 4$. Here these divergences manifest
themselves as poles in the appearing gamma functions.

In order to isolate these poles, we expand the expressions
above around $n=4$. We immediately add the two integrals
and obtain for the pole term $P$:
\begin{equation}
P=\frac{i\pi^2}{3(2 \pi)^4} \frac{1}{4-n} 
\left[ p^\mu p^\nu-p^2 g^{\mu \nu} \right] \, .
\end{equation}
As a consequence of the dimensional
regularization scheme, it is gauge invariant. Neglecting
$O(4-n)$ yields the finite contribution $F$,
\begin{eqnarray}
F &=& \frac{i\pi^2}{(2 \pi)^4} \left(\frac{1}{6}\gamma_E + \frac{1}{2}
\int_0^1 dx \, (1-2x)^2 \ln\left(\frac{f}{4\pi M_0^2}\right)\right)
\left[ p^\mu p^\nu-p^2 g^{\mu \nu} \right]  \nonumber \\
&+& \frac{i\pi^2}{(2 \pi)^4} \left[ \frac{p^2}{6} -
m^2\ln\left(\frac{m^2}{4\pi M_0^2}\right) +
\int_0^1 dx \, \left(\frac{1}{2}p^2
(1-2x)^2 + f\right)\ln\left(\frac{f}{4 \pi M_0^2}\right) \right] g^{\mu \nu} \,.
\end{eqnarray}
We separated the gauge invariant term. Adding the pole term given
above to it, yields the result presented in the main text.

Thus it remains to be shown that the second, gauge variant, contribution
$\propto g^{\mu \nu}$ vanishes. In order to get rid of the
logarithmic function one integrates by parts and finds
\begin{equation}
\left[....\right] = p^2 \left[\frac{1}{6} -\int_0^1 dx\, \frac{x(2x-1)}{f}
\left(m^2+p^2(x^2-\frac{3}{2}x+\frac{1}{2})\right) \right]\, .
\end{equation}
Note that this expression is independent on the reference mass.
We substitute $z=2x-1$ and indeed obtain for the integral
\begin{eqnarray}
\int_0^1 dx\, .... &=& \frac{1}{2}\int_{-1}^1 dz \,
\frac{\frac{1}{2}z(z+1)}{m^2-\frac{1}{4}p^2(1-z^2)-i\epsilon} 
\left(m^2+\frac{1}{4}p^2(z^2-z)\right) \nonumber \\
%&=& \frac{1}{4}\int_{-1}^1 dz \,
%\frac{1}{m^2-\frac{1}{4}p^2(1-z^2)} 
%\left(z m^2+ z^2(m^2 -\frac{1}{4}p^2(1-z^2))\right) \nonumber \\
&=& \frac{1}{4}\int_{-1}^1 dz \, z^2 = \frac{1}{6} \,.
\end{eqnarray}

% PHYSREV references
%\begin{thebibliography}{99}
%\bibitem{Sie} A. J. F. Siegert, Phys. Rev. {\bf 52}, 787 (1937).
%\bibitem{Naus} H. W. L. Naus, Phys. Rev. C {\bf 55}, 1580 (1997).
%\bibitem{Hy} H. Hyuga and H. Ohtsubo, Nucl. Phys. {\bf A294}, 348 (1978).
%\bibitem{FSK} N. Fukuda, K. Sawada and M. Taketani, 
%Prog. Theor. Phys. {\bf 12}, 156 (1954);
%S. Okubo, Prog. Theor. Phys. {\bf 12}, 102; 603 (1954).
%\bibitem{Sato} T. Sato and T.-S. H. Lee, Phys. Rev. C {\bf 54}, 2660 (1996).
%\bibitem{Adam} J. Adam, Jr., E. Truhl\'{i}k and D. Adamov\'{a},
%Nucl. Phys. {\bf A492}, 556 (1989);
%J. Adam, Jr., Ch. Hajduk, H. Henning, P. U. Sauer and E. Truhl\'{i}k,
%Nucl. Phys. {\bf A531}, 623 (1991); J. Adam, Jr., H. G\"{o}ller
%and H. Arenh\"{o}vel, Phys. Rev. C {\bf 48}, 370 (1993).
%\bibitem{Friar} J. L. Friar, Ann. Phys. (N.Y.) {\bf 104}, 380 (1977).
%\bibitem{WT} J. C. Ward, Phys. Rev. {\bf 78}, 182 (1950);
%Y. Takahashi, Nuovo Cimento {\bf 6}, 371 (1957).
%\bibitem{Ry} L. H. Ryder, {\it Quantum Field Theory, Second edition}
%(Cambridge University Press, 1996).                                           
%\bibitem{Schw} J. Schwinger, Phys. Rev. {\bf 82}, 664 (1951).
%\bibitem{BD} J. D. Bjorken and S. D. Drell, {\it Relativistic Quantum
%Mechanics} (MacGraw-Hill, New York, 1964).
%\bibitem{Nabo} H. W. L. Naus, J. W. Bos and J. H. Koch, 
%Int. J. Mod. Phys. A {\bf 7}, 1215 (1992).
%\end{thebibliography}

%NUCPHYS REFERENCES

\end{document}